\documentclass[preprint]{ptephy_v1}

\preprintnumber{XXXX-XXXX} 
\usepackage{hyperref}

\begin{document}

\title{CALPAGAN: Calorimetry for Particles using GANs}
\author{Ebru Simsek}
\affil{Physics Department, Bogazici University}

\author{Bora Isildak}
\affil{Physics Department, Yildiz Technical University}

\author{Anil Dogru \& Reyhan Aydogan }
\affil{Department of Computer Science, Ozyegin University}

\author{Burak Bayrak \& Seyda Ertekin}
\affil{Department Of Computer Engineering, Middle East Technical University}


\begin{abstract}
In this study, a novel approach is demonstrated for converting calorimeter images from fast simulations to those akin to comprehensive full simulations, utilizing conditional Generative Adversarial Networks (GANs). The concept of Pix2pix is tailored for CALPAGAN, where images from fast simulations serve as the basis(condition) for generating outputs that closely resemble those from detailed simulations. The findings indicate a strong correlation between the generated images and those from full simulations, especially in terms of key observables like jet transverse momentum distribution, jet mass, jet subjettiness, and jet girth. Additionally, the paper explores the efficacy of this method and its intrinsic limitations. This research marks a significant step towards exploring more efficient simulation methodologies in High Energy Particle Physics.

\end{abstract}

\subjectindex{Machine Learning, High Energy Physics}

\maketitle

\section{Introduction}
In the field of experimental particle physics, comprehensive simulations are made to estimate what the obtained experimental data will look like in the context of the Standard Model. These simulations begin with the initial hard collision and extend to the generation of electronic signals in the detector. The most computationally demanding aspects involve modeling the detector and the detailed step-by-step simulation of how particles interact with it, particularly within the calorimeter. Notably, the Geant4 toolkit \cite{geant4}, which includes cutting-edge models, is used for simulating particle detectors at CERN's LHC. 

The high energy physics community, particularly with the advent of the High-Luminosity Large Hadron Collider (HL-LHC), has been increasingly engaged in the application of deep learning techniques, with a specific focus on Generative Adversarial Networks (GANs) \cite{musella}. Introduced by Goodfellow et al. in 2014 \cite{Goodfellow}, GANs comprise two deep neural networks: the generator and the discriminator. The move towards utilizing Generative Adversarial Networks (GANs) is mainly driven by the intense computational requirements linked with the High-Luminosity Large Hadron Collider (HL-LHC). This is particularly relevant in the areas of detector modeling and the comprehensive simulation of particle interactions in the calorimeter, tasks that are recognized as some of the most demanding in terms of CPU usage.\cite{CPU}. The HL-LHC's requirements for fast and accurate simulation methods are critical, as over half of the computational resources for Large Hadron Collider experiments are dedicated to these simulations \cite{Giammanco:2014bza} \cite{hep}.

In this study, focus shifts from the traditional full simulation approach, known for its high CPU consumption, to exploring the capabilities and limitations of a deep learning-based GAN. This is aimed at achieving both high precision and rapid calorimeter simulations. The goal is to develop an innovative fast simulation technique that could significantly reduce both computation time and disk space requirements in LHC and future physics research. \cite{paganini}

The structure of this paper is organized as follows: Section \ref{sec:Calorimeter data simulation and input data pre-processing} offers a comprehensive description of the simulation and reconstruction processes for the input images (condition) as well as the target images. Section \ref{sec:network_structure} details the architecture of both the generator and discriminator networks. Subsequently, Section \ref{sec:results} presents a performance evaluation of the generated outcomes by comparing the distributions of high-level jet properties.

\newpage
\section{Calorimeter data simulation and input data pre-processing}
\label{sec:Calorimeter data simulation and input data pre-processing}
Monte Carlo datasets were generated for CMS Detector by utilizing CMSSW\cite{cmssw}, the software framework designed for CMS. CALPAGAN needs a Delphes image as input along with a noise, and the target image is the image obtained from GEANT4 simulation which is only possible using CMSSW.  This extensive procedure encompassed several stages: Generation, Simulation, Digitization, and Reconstruction, collectively known as GEN-SIM-DIGI-RECO chain. The event simulation was carried out using both Delphes and Geant4. 

\begin{figure}[h!]
    \centering
    \includegraphics[width=0.4\linewidth]{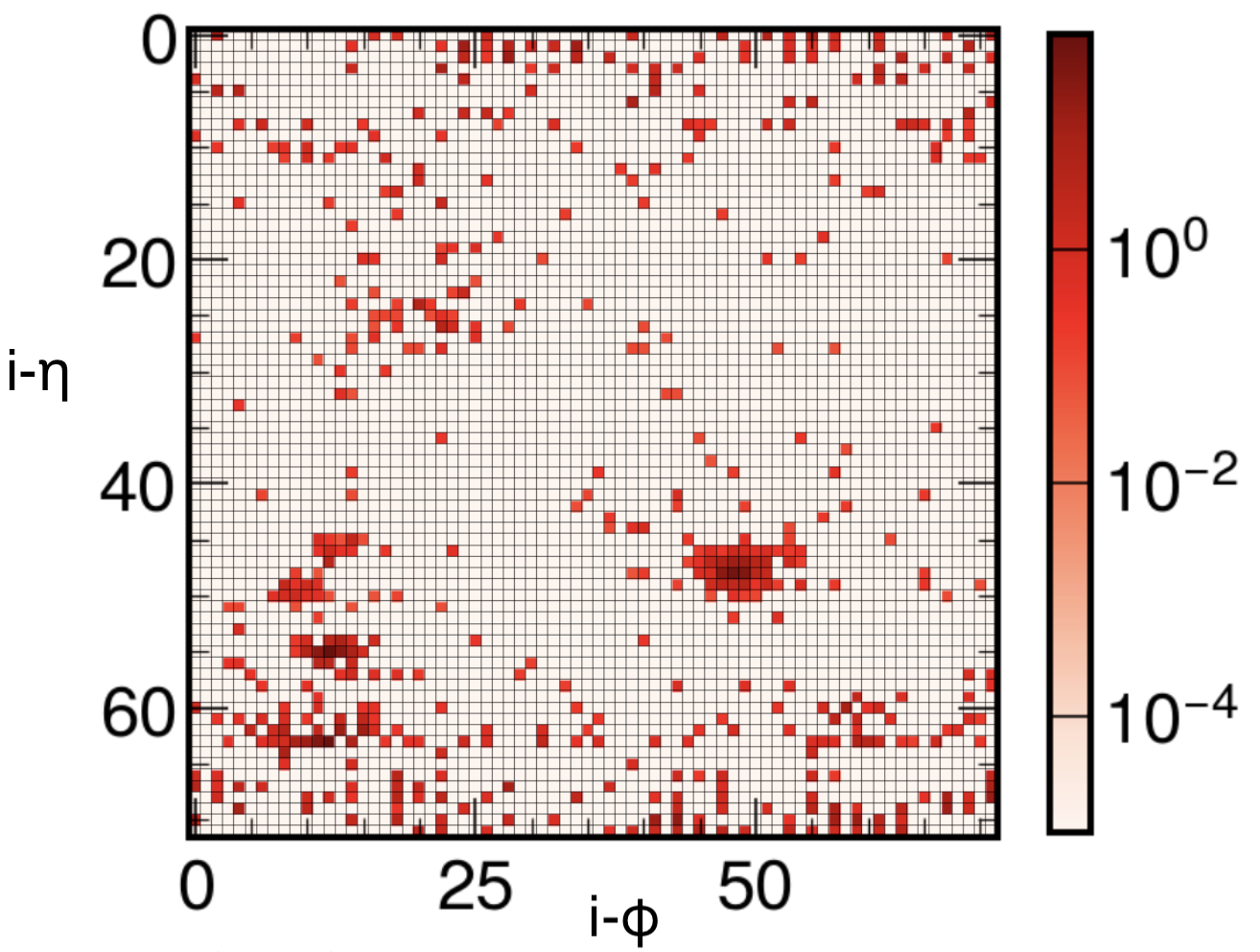}
    \caption{This image displays the calorimetric energy in GeV of an event created with Geant4, presented on a 72 x 72 grid (5184 pixels total) within the $\eta$-$\phi$ space. 
    Here, the horizontal axis represents the  $\eta$ indices and the vertical axis represents the  $\phi$  indices respectively.  }
    \label{fig:Event}
\end{figure}

In this research, Geant4 and Madgraph4 \cite{madgraph4} combined with Pythia8 and Delphes \cite{delphes} were independently used to generate 10,000 events each of W+Jet (with up to 3 jets) and dijet. To ensure the generation of identical events, both simulations utilized the same Pythia8 output, which was stored in “HepMC” format files. The dataset was divided into two equal parts: 5,000 events for training and 5,000 for testing. Subsequently, two-dimensional calorimeter images were produced by summing the hadronic and electromagnetic energy in the calorimeter towers. The calorimeter structure of the CMS Detector allows the creation of a $72\times 72$ image in $\eta-\phi$ space. Moreover, since the pixel values in these images represent energy rather than color (as shown in Fig. \ref{fig:Event}), they are single-channel. This characteristic means that the standard Pix2pix\cite{pix2pix} GAN architecture cannot be directly applied, thus necessitating some modifications to the network structure. Pix2pix is successful in creating synthetic images that closely resemble real images. For example, when tasked with creating artificial tree images, a GAN based on Pix2pix outperforms a GAN designed to mimic images created by particles dispersing energy at specific points in the calorimeter.

\begin{figure}[h!]
    \centering
    \includegraphics[width=0.9\linewidth]{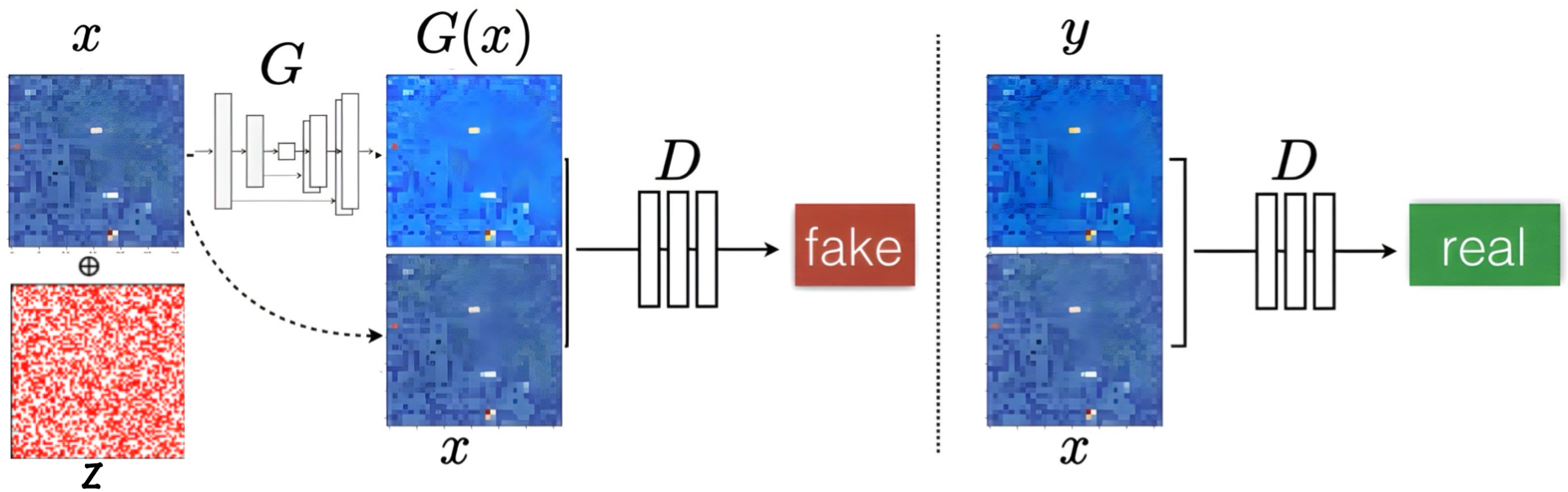}
    \caption{Training architecture of CALPAGAN: x represents the input images, which are calorimeter images produced by Delphes. z is random noise. G(x) denotes the generated data, while y signifies the real images which is calorimeter images produced by Geant4. D represents the discriminator.}
    \label{fig:GAN_new}
\end{figure}

As seen in Fig. \ref{fig:GAN_new}, z, the random noise, has been added to calorimeter images produced by Delphes,  x, then served as input for the generator G. The generated data G(x) is then added pixel by pixel to x, producing a combined input G(x)+x for the discriminator D. The discriminator evaluates both this combined input and the real samples y, learning to distinguish between real and counterfeit data through iterative training.

The generator of CALPAGAN creates random distributions using the calorimetric jet data from Delphes. Finally, the discriminator network (D) decides how realistic the distribution produced by the generator has produced thanks to simulations performed by Geant4 as illustrated in Fig. \ref{fig:GAN_new}. 

Actually, this is a game played by the generator and the discriminator, known as a zero-sum game \cite{Nash} in the literature.The generator network creates data and engages in a zero-sum game with a discriminator network, aiming to reach a Nash equilibrium \cite{Nash} based on the feedback from the discriminator. At equilibrium, the discriminator must randomly guess with a 50 chance whether the generator's data is real or fake. 

\section{Network Structure and Method}
\label{sec:network_structure}
\subsection{The CALPAGAN}
\label{sec:Calpagan}
The model employed in this study, as shown in Fig., known as cGAN (conditional GAN)\cite{cGAN}, is designed to transform the input image, which also serves as the condition, into the target image.

\begin{figure}[h!]
    \centering
    \includegraphics[width=0.7\textwidth]{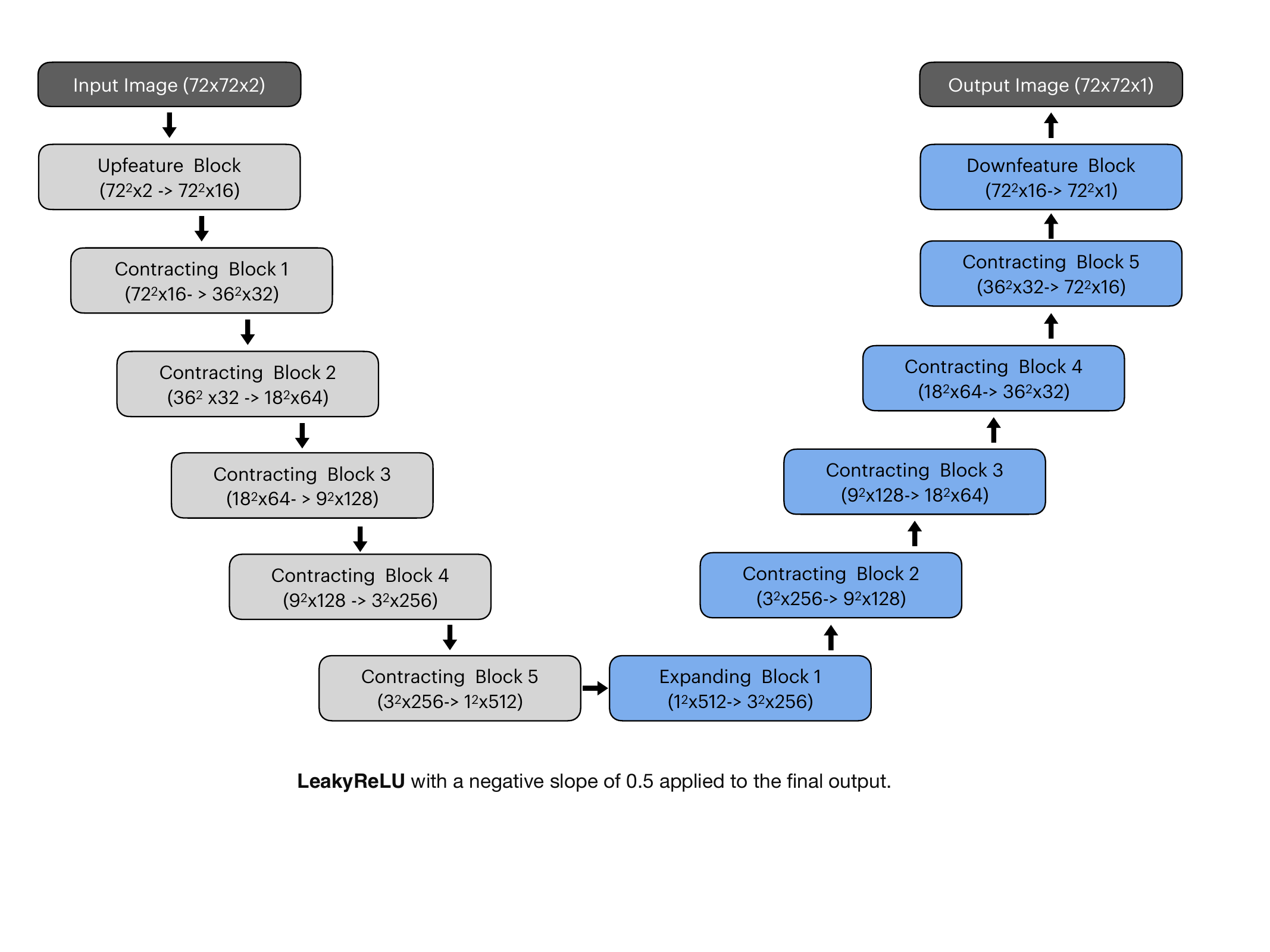}
    \caption{The U-Net architecture of the generator of CALPAGAN.}
    \label{fig:Unet}
\end{figure}

In the Pix2pix framework, the generator's architecture is based on U-Net\cite{Olaf}. U-Net is an advanced deep learning framework that utilizes an encoder-decoder structure with convolutional networks to efficiently process and analyze data. Its unique feature, the incorporation of skip connections, ensures the preservation of important spatial information. As illustrated in Fig. \ref{fig:Unet}, U-Net structure first compresses the input image down to a bottleneck size, and then, following the bottleneck, it expands the image back to the required output size.

The generator was designed using the U-Net architecture, incorporating five contracting blocks in its encoder network. Every contracting block in the sequence includes the following computational steps: an initial convolutional layer (with a kernel size of 3, padding of 1, stride of 1, and a doubling of the number of input channels to generate the output channels), followed by batch normalization, then the Leaky ReLU (Rectified Linear Unit) activation function (having a negative slope of 0.2). This is succeeded by a second convolutional layer (employing the same hyperparameters as the first), another round of batch normalization, a further application of the Leaky ReLU activation (also with a negative slope of 0.2), and concluding with a max pooling layer at the block's end. It's important to mention that in the first three contracting blocks, a max pooling layer is used with a kernel size of 2 and a stride of 2. However, in the final two blocks, the max pooling layer's hyperparameters change to a kernel size of 3 and a stride of 3. Additionally, at both the start and finish of the generator, there is a feature map block in use, which includes a single convolutional layer with a kernel size of 1. Initially, this block converts the input image channels into 16 channels, and ultimately, it diminishes the channel count to just 1.

 During the decoding stage, five expanding blocks are employed, each serving as a reverse counterpart to the five contracting blocks in the encoding phase. Each expanding block structured as follows: it begins with a bilinear upsampling layer, followed by a first convolutional layer (with a kernel size of 2, no padding, stride 1, and the number of output channels being half that of the input channels). Next is a second convolutional layer (with a kernel size of 3, padding 1, stride 1, and halving the number of input channels), which receives inputs from both the skip-connection and the output of the first convolutional layer. This is followed by batch normalization and the Leaky ReLU activation function (with a negative slope of 0.2). A third convolutional layer (with kernel size 2, padding 1, stride 1, and matching the number of input channels with the output channels), another round of batch normalization, and a second application of the Leaky ReLU activation function (also with a negative slope of 0.2) follow. It's important to note that the scaling factor of the upsampling layer is set to 3 for the first two layers and 2 for the rest.

The discriminator employs a feature map block to convert the input image into an 8-channel image. This is followed by the use of three sequential contracting blocks for additional data processing. Subsequently, a final feature map block is used to transform the input into a single-channel image. The feature map and contracting blocks in the discriminator mirror those in the generator. However, the discriminator distinguishes itself by favoring instance normalization over batch normalization in its contracting blocks.

Additionally, the Delphes calorimeter images used in the training process were normalized by dividing by the highest energy pixel value of each Delphes image, and similarly, the Geant4 images were normalized by the highest energy pixel value of each Geant4 image. This normalization moved pixel values to the range [0, 1], which also partially addressed the sparsity problem by eliminating the impact of zero-energy pixels on the loss function (as opposed to when pixel values were previously normalized to the range [-1, 1], where zero-energy pixels would contribute a value of -1 to the loss function, leading to the generator producing blank images after a few epochs.). 

The loss function for the CALPAGAN is composed of two distinct components. The first is the conventional loss function typically employed in cGANs. The second component is the L1 loss function, also known as the mean absolute error (MAE) loss function. This is calculated as the average of the absolute differences and is particularly effective in reducing blurriness in the images as shown in below. By integrating these two loss functions, we formulated the following loss function \ref{loss_formula} , which was, of course, aimed to be minimized during the training process. In the formula, x, y, and z represent the input image, output image, and random noise, respectively.

\begin{equation}
\begin{aligned}
    \mathcal{L}_{cGAN}(G,D) &= \mathbb{E}_{x,y}[\log D(x,y)] + \mathbb{E}_{x,z}[\log(1 - D(x, G(x,z)))] \\
    \mathcal{L}_{L1}(G) &= \mathbb{E}_{x,y,z}[\|y - G(x,z)\|_{1}] \\
    G^{*} &= \arg \min_{G} \max_{D} \mathcal{L}_{cGAN}(G, D) + \lambda \mathcal{L}_{L1}(G)
\end{aligned}
\label{loss_formula}
\end{equation}


The jets were detected and identified using the FastJet program \cite{fastjet}, which utilizes the electromagnetic and hadronic energy depositions in the calorimeter(ECAL+HCAL) towers and the outcomes from the Geant4 simulation. Subsequently, a trimming process, as described by Krohn et al. in 2010 \cite{jettrimming}, was employed to remove energy contributions from pile-up events.

\begin{figure}[h!]
    \centering
    \includegraphics[width=0.9\linewidth]{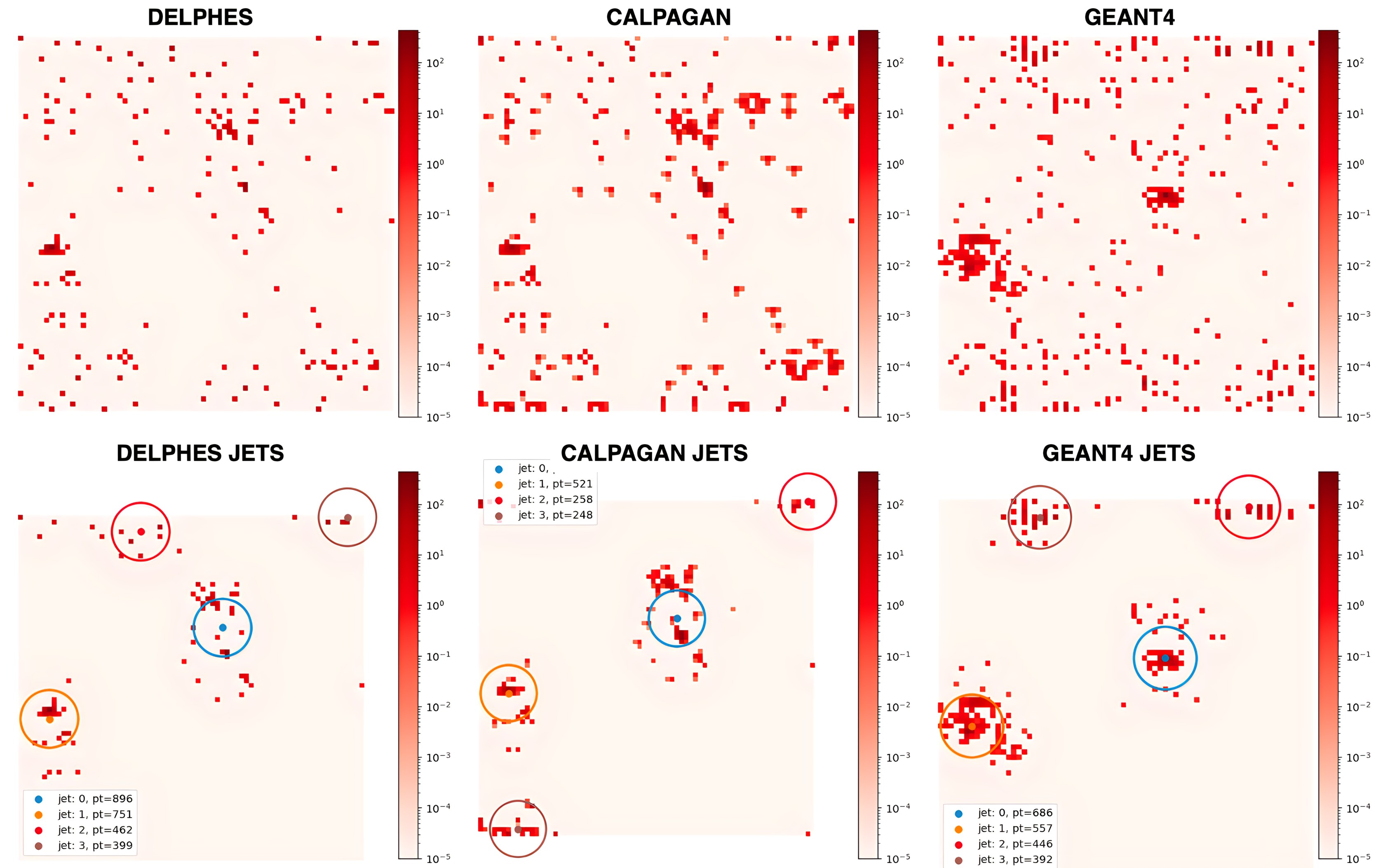}
    \caption{Comparison of event images obtained from the test set with Delphes and Geant4 event images (top: a randomly selected example from 5000 test events). The bottom image shows the images of jets obtained from these event images in the  $\eta$ (horizontal) and $\Phi$ (vertical) space.}
    \label{fig:jet_comparison}
\end{figure}

\newpage
\section{Results and Conclusion}
\label{sec:results}
 We first tried to apply Pix2pix architecture as it is. However, it didn’t perform so well. So, we have tweaked the architecture to get better results. Inspired by the Pix2pix, CALPAGAN has been developed with the aim of creating calorimeter images similar to those produced by Geant4, using Delphes images as a condition. CALPAGAN has been effective in producing images very close to those from Geant4, thereby reducing computational costs. However, during this research, various challenges were encountered and the most significant of which was the natural sparsity observed in the data.

Data sparsity limits the success of CALPAGAN. The presence of numerous zero-valued pixels in the calorimeter images adversely affected weight calculations. Despite testing with various activation functions such as ReLU, sigmoid, and softmax, as well as experimenting different filter sizes and strides within the generator network, these attempts failed to yield positive results.

Nonetheless, visually identifying the similarities between synthetic calorimeter images and their real counterparts remains a complex task. The widely recognized metric for evaluating GANs is the Fréchet Inception Distance (FID)\cite{FID}. Consequently, FID was employed to assess the similarity between the generated and real images.



\begin{figure}[h!]
    \centering
    \includegraphics[width=1.0\linewidth]{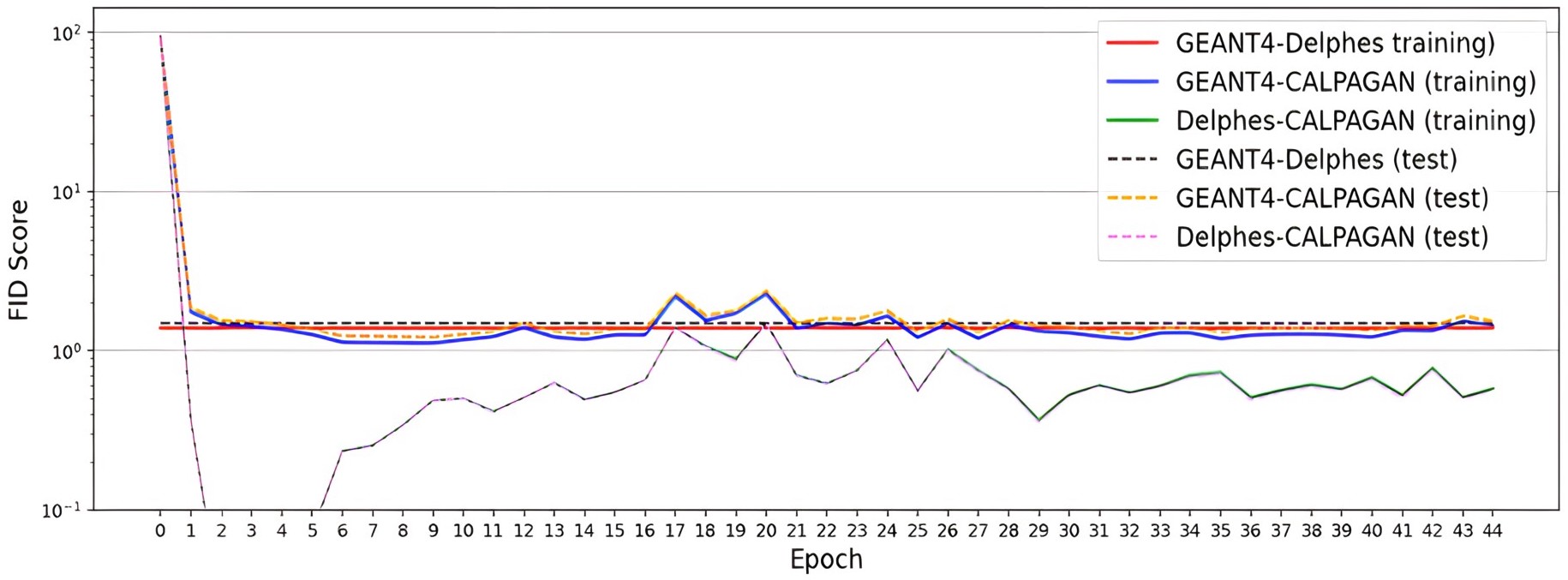}
    \caption{The evolving FID score with increasing number of epoch.}
    \label{fig:fid_score}
\end{figure}

Throughout the training phase, the Wasserstein-2 distance or Fréchet Inception Distance (FID) score between these three image sets is presented in Fig.\ref{fig:fid_score} across various epochs. It is noticeable that beyond a specific epoch, the anticipated relationship is attained. This suggests that in terms of the FID score, the calorimeter images produced by CALPAGAN are situated in a proximity closer to the Geant4 images as opposed to the Delphes images. 

 \newpage
However, this closeness is valid in training for a particular physical process. Training with W+Jet events gave the desired results for the W+Jet test sample, while training with dijet events gave the desired results for the dijet test sample. However, neither the training performed with W+Jet events was successful in giving the desired results for the dijet test sample nor the training performed with the dijet events was successful in giving the desired results for the W+Jet test sample.

\begin{figure}[h!]
    \centering
    \includegraphics[width=\linewidth]{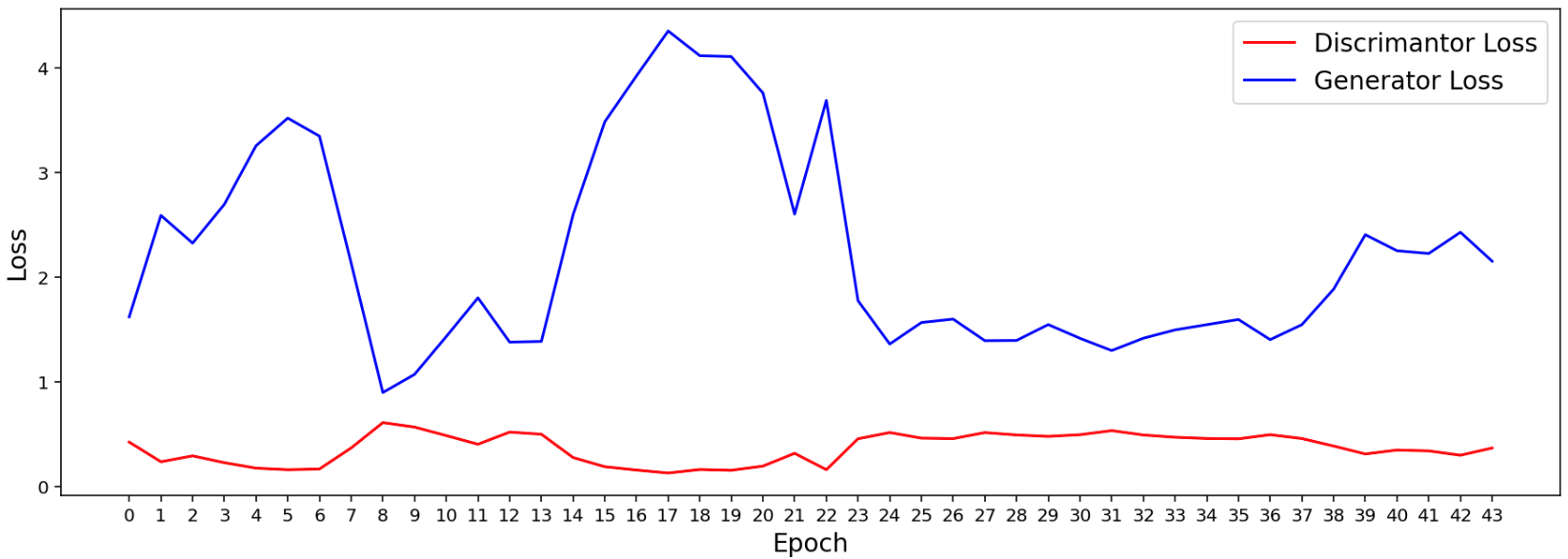}
    \caption{The change in the loss function with the number of epochs during training with jet images.}
    \label{fig:fid_score_jets}
\end{figure}

Additonally, the Fig. \ref{fig:fid_score_jets} shows the change in the loss function values obtained from the test sample in the GAN training conducted with jet images as the number of epochs progresses.

The loss value of the Generator in a CALPAGAN remains continuously high, and the image generation performance is relatively low, it typically suggests that the Generator is struggling to produce realistic images that can fool the Discriminator.

In our CALPAGAN model, inspired by the Pix2pix architecture, we observed that the discriminator learns rapidly, outpacing the generator and resulting in unstable training dynamics. This quick learning provides overly strong gradient feedback, making it difficult for the generator to catch up. Additionally, the architectural imbalance between the simpler discriminator and the more complex generator makes the problem worse, leading to poor performance. This issue can also stem from the natural sparsity of the data. Consequently, designing a GAN to mimic images of energy dispersion by particles at specific locations within a calorimeter becomes challenging.

The red line represents the loss of the discriminator. It is much steadier and remains low, which implies that the discriminator is performing its task with relative ease. The low and steady loss suggests that the discriminator quickly becomes good at distinguishing real images from the fake ones generated by the generator.

Therefore the generator might need a more complex architecture to improve its performance. here might be a need for techniques to stabilize the training, such as modified loss functions, regularization methods, or different training strategies which we have already tried some of them.

\begin{figure}[h!]
    \centering
    \includegraphics[width=0.75\linewidth]{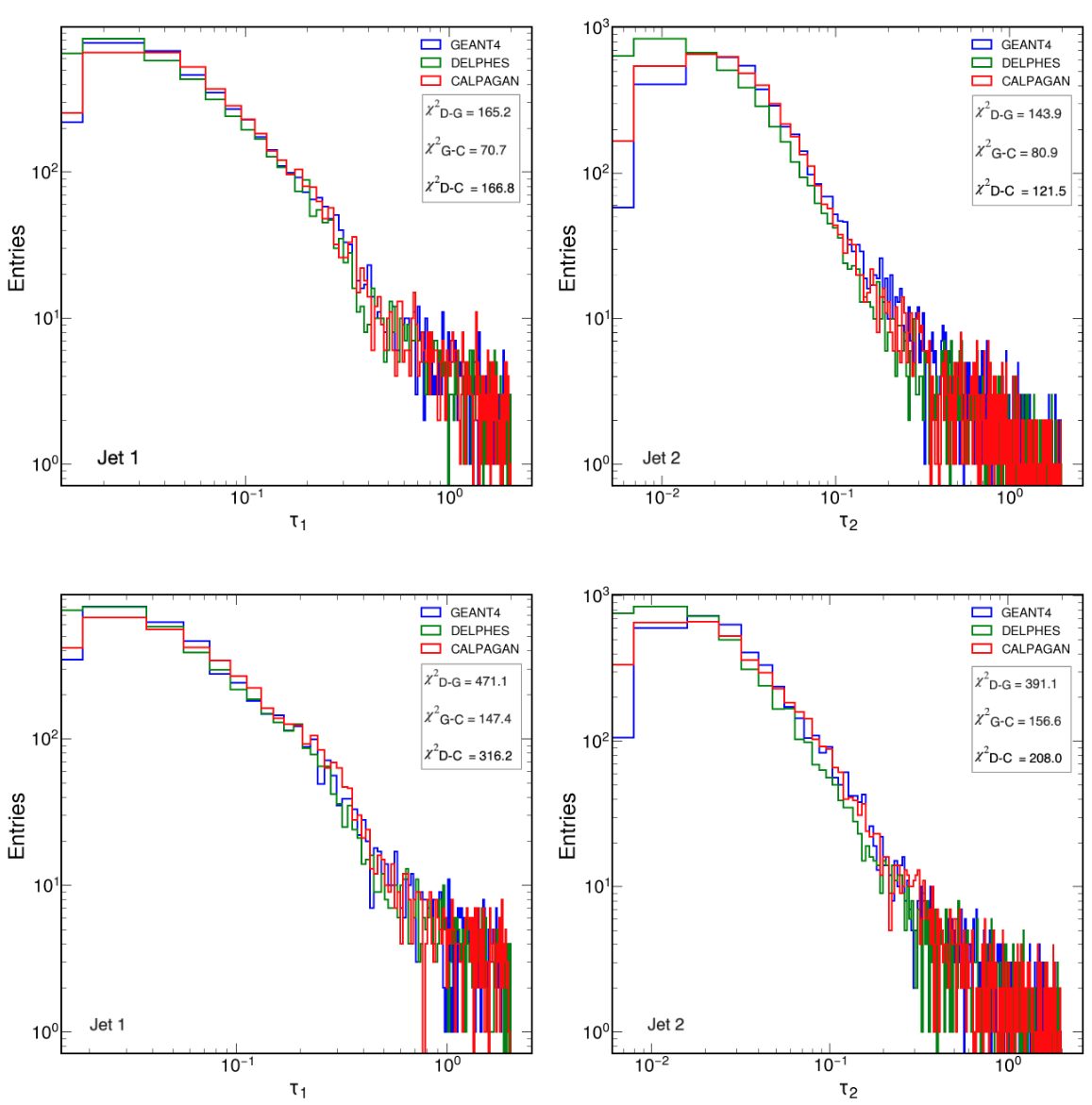}
    \caption{The comparison of the 1-subjettiness  $\tau_1$ (top) and 2-subjettiness  $\tau_2$ (bottom) distributions obtained from Geant4, Delphes, and CALPAGAN for the first and second jets.For each comparison, the $\chi^2$ distances between Delphes-Geant4 (D-G), Geant4-CALPAGAN (G-C), and Delphes-CALPAGAN (D-C) distributions are provided.}
    \label{fig:subjettiness}
\end{figure}

In Fig.\ref{fig:subjettiness}, the distributions of 1-subjettiness $\tau_1$ and 2-subjettiness $\tau_2$ obtained from Geant4, Delphes, and CALPAGAN are shown for the first and second jets.'N-subjettiness-- designed to identify boosted hadronically-decaying objects like electroweak bosons and top quarks. Combined with a jet invariant mass cut, N-subjettiness is an effective discriminating variable for tagging boosted objects and rejecting the background of QCD jets with large invariant mass' \cite{subjettiness}, is particularly suitable for the Delphes-Geant4-CALPAGAN comparison due to its variable nature that captures the internal structure of jets.




\newpage

The linear radial moment or girth is a specific case where a general radial moment is defined as f(r)=r \cite{girth}. It is defined based on the components that constitute a jet as follows: 
\[g =\sum_{i \in jet} \frac{(p^i_T)}{p^{jet}_T}{r_i}\]

   \begin{figure}[h!]
        \centering
        \includegraphics[width=0.18\linewidth]{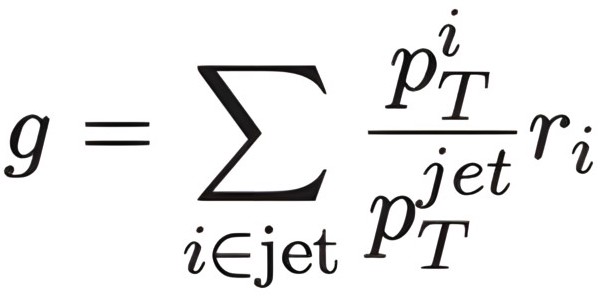}
   \end{figure}

\begin{figure}[h!]
    \centering
    \includegraphics[width=\linewidth]{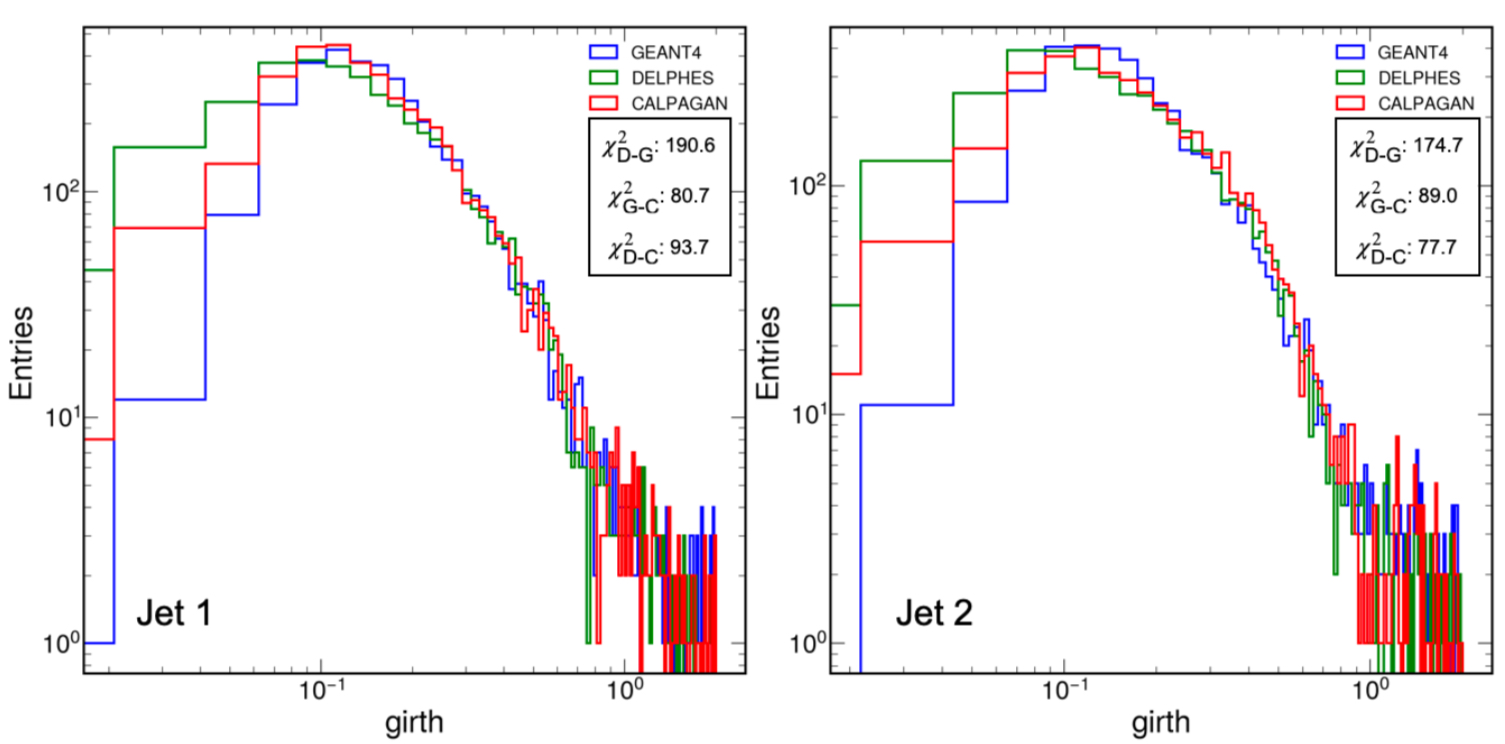}
    \caption{ Comparison of the girth distributions obtained from Geant4, Delphes, and CALPAGAN for the first and second jets. }
    \label{fig:girth}\end{figure}

Comparison of the girth distributions obtained from Geant4, Delphes, and CALPAGAN for the first and second jets is shown in Fig. \ref{fig:girth}

\begin{figure}[h!]
    \centering
    \includegraphics[width=0.9\textwidth]{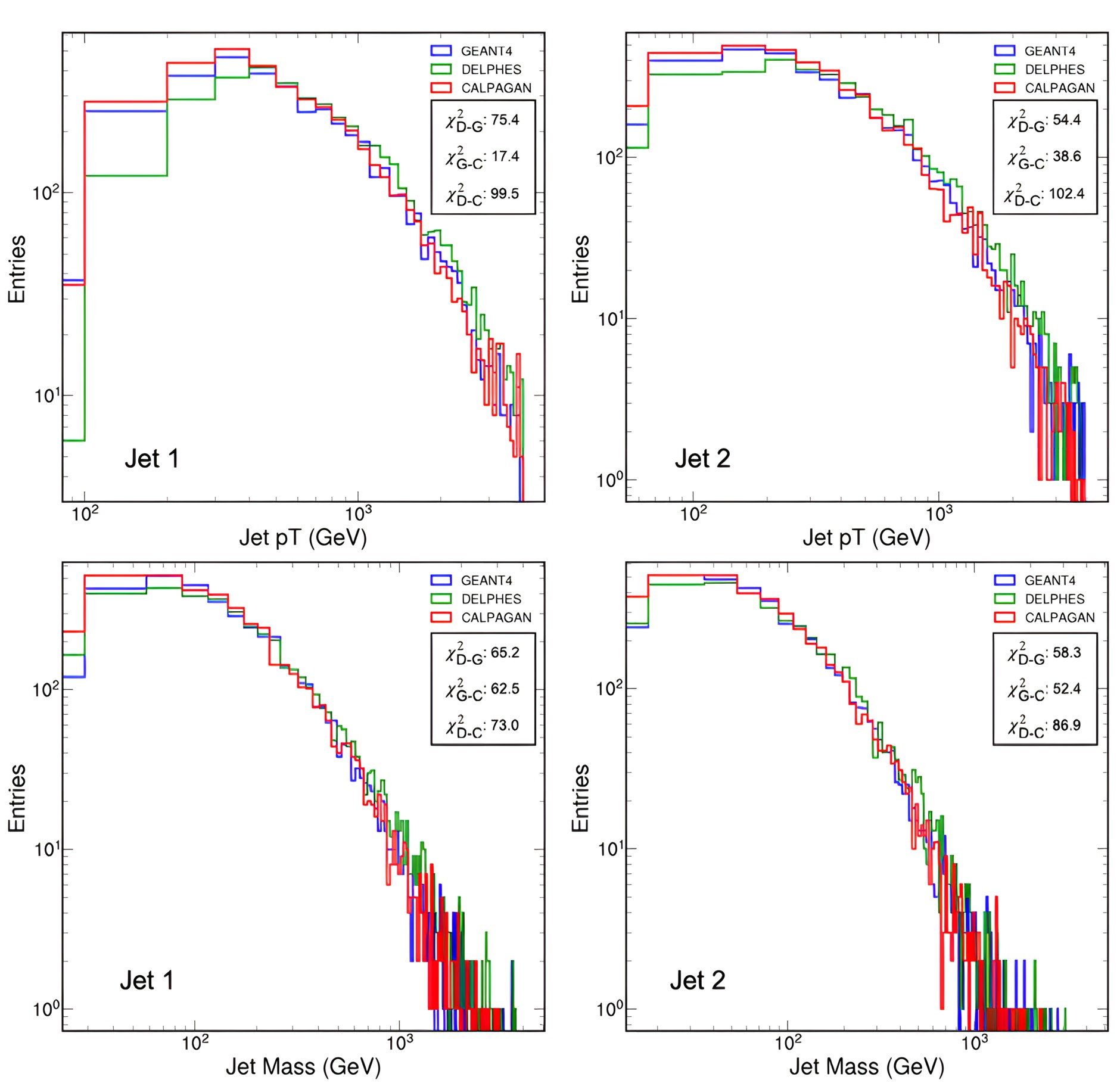}
    \caption{The comparison is presented for the distributions of transverse jet momentum (top) and jet mass (bottom) obtained from Geant4, Delphes, and CALPAGAN for the first and second jets.}
    \label{fig:jet_mass_pt}
\end{figure}

In Fig.\ref{fig:jet_mass_pt}, the comparison is presented for the distributions of transverse jet momentum (top) and jet mass (bottom) obtained from Geant4, Delphes, and CALPAGAN for the first and second jets. 

\begin{figure}[h!]
    \centering
    \includegraphics[width=\textwidth]{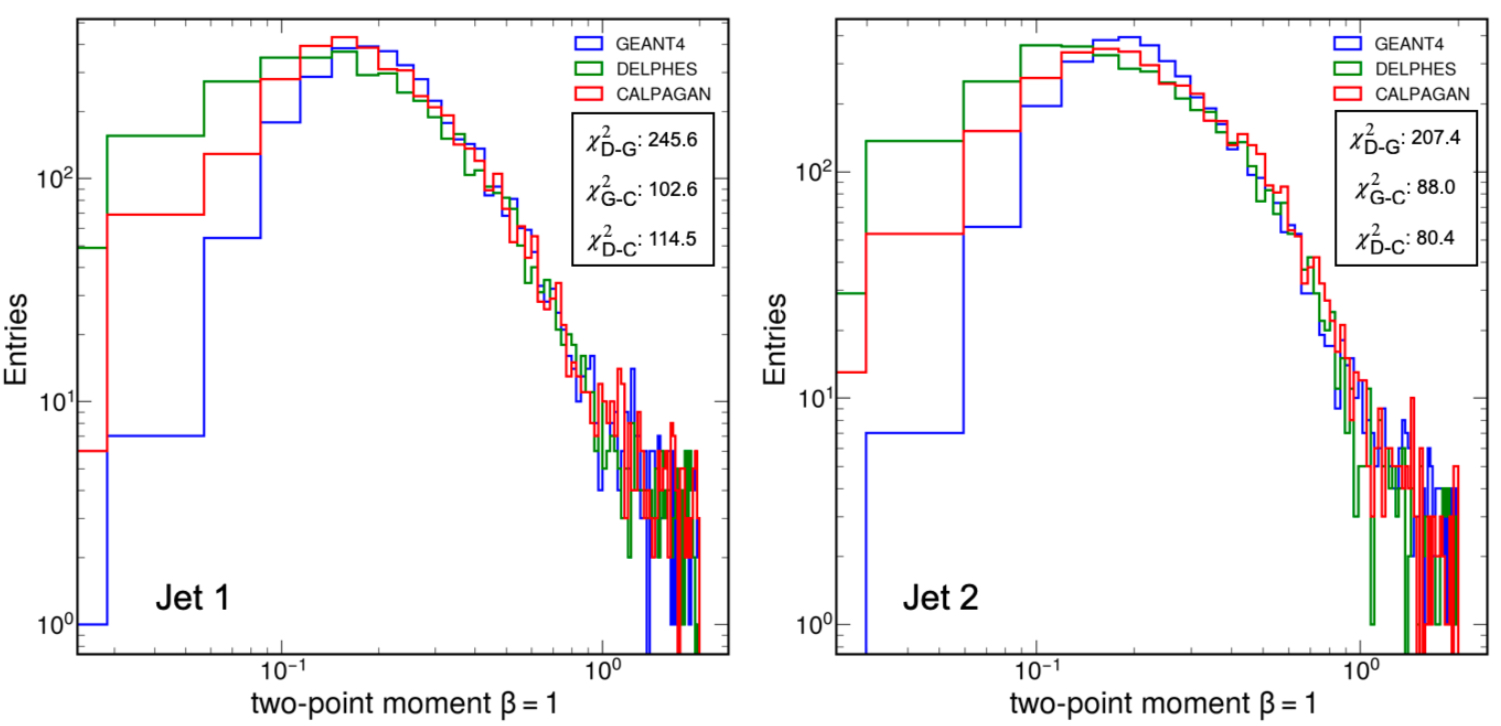}
    \caption{Comparison of the two-point moment distributions obtained from Geant4, Delphes, and CALPAGAN for the first and second jets }
    \label{2_point_moment}
\end{figure}

\newpage
Comparison of two-point moment distributions obtained from Geant4, Delphes, and CALPAGAN for the first and second jets can be seen in Fig.\ref{2_point_moment}.

\begin{figure}[h!]
    \centering
    \includegraphics[width=\textwidth]{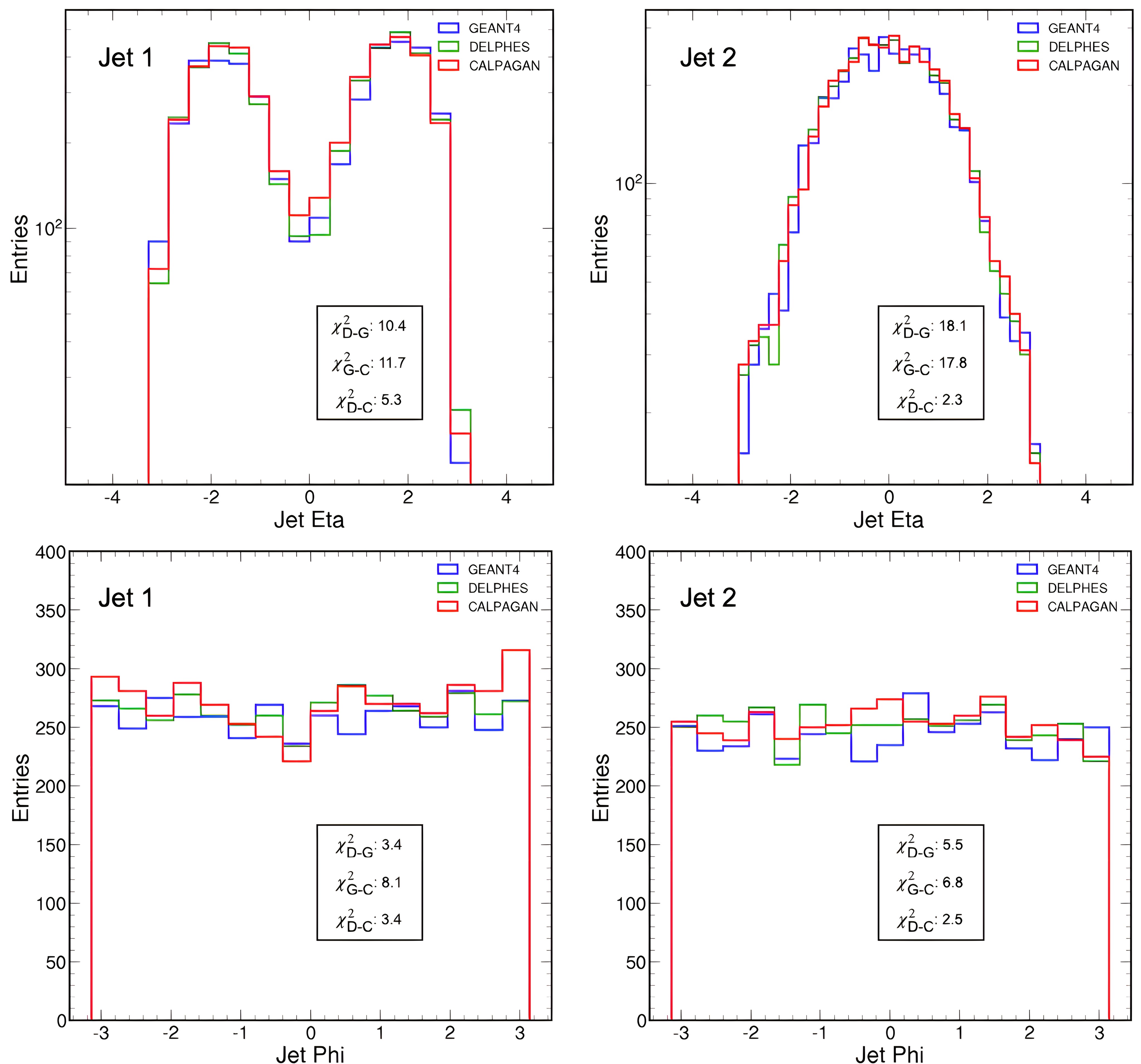}
    \caption{Comparison of transverse jet pseudorapidity $\eta$ and jet azimuthal angle $\Phi$ distributions obtained from Geant4, Delphes, and CALPAGAN for the first and second jets.}
    \label{fig:jet_eta_phi.dist.jpg}
    \end{figure}

\newpage
The Fig.\ref{fig:jet_eta_phi.dist.jpg} depicts the distributions of the differences in pseudorapidity $\eta$  and azimuthal angle $\Phi$ between the first and second jets are compared for events obtained from Geant4, Delphes, and CALPAGAN.

\newpage

\begin{figure}[h!]
    \centering
    \includegraphics[width=0.8\linewidth]{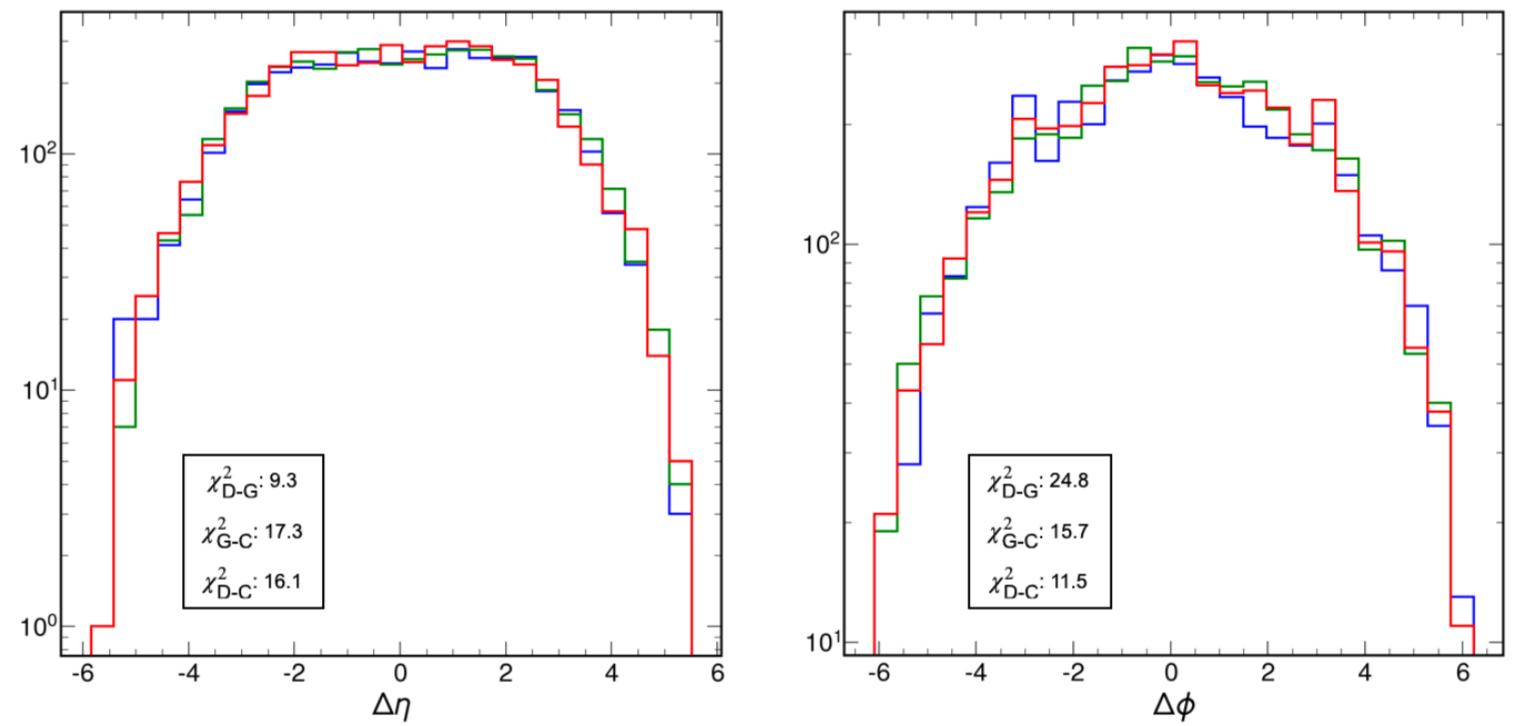}
    \caption{The distributions of the differences in pseudorapidity $\eta$ and azimuthal angle $\Phi$ between the first and second jets are compared for events obtained from Geant4, Delphes, and CALPAGAN.}
    \label{fig:jet_pseudo_dist}
\end{figure}

The Fig. \ref{fig:jet_pseudo_dist} displays the distributions of the differences in pseudorapidity $\eta$ and azimuthal angle $\Phi$ between the first and second jets are compared for events obtained from Geant4, Delphes, and CALPAGAN.
 
\newpage
Table \ref{table_1} displays the $\chi^2$ distances between the distributions of various jet variables obtained from Geant4, Delphes, and CALPAGAN for W+Jet events. 

\begin{table}[h!]
\centering
\caption{The $\chi^2$ distances between the distributions of various jet variables obtained from Geant4, Delphes, and CALPAGAN for W+Jet events.}
\label{table_1}
\begin{tabular}[H]{|c|c|c|c|c|c|c|} 
\hline & \multicolumn{6}{|c|}{$\chi^2$ distances } \\
\hline & \multicolumn{2}{|c|}{ Delphes-Geant4 } & \multicolumn{2}{|c|}{ Geant4-CALPAGAN } & \multicolumn{2}{|c|}{ Delphes-CALPAGAN } \\
\hline & Jet 1 & Jet 2 & Jet 1 & Jet 2 & Jet 1 & Jet 2 \\
\hline 1-subjettiness & 179.0 & 140.3 & 137.0 & 94.5 & 245.9 & 88.1 \\
\hline 2-subjettiness & 431.0 & 400.3 & 158.4 & 242.2 & 342.6 & 147.5 \\
\hline Jet girth & 190.6 & 174.7 & 172.3 & 77.7 & 302.4 & 84.3 \\
\hline 2 point moment & 245.6 & 207.4 & 132.5 & 85.1 & 310.5 & 86.2 \\
\hline Jet $\mathbf{p}_{\mathrm{T}}$ & 75.4 & 54.4 & 62.8 & 44.4 & 207.3 & 89.3 \\
\hline Jet mass & 65.2 & 58.3 & 47.9 & 44.8 & 54.7 & 55.3 \\
\hline Jet $\eta$ & 10.4 & 18.1 & 19.0 & 20.5 & 6.5 & 13.2 \\
\hline Jet $\phi$ & 3.4 & 5.5 & 3.0 & 7.4 & 1.2 & 4.1 \\
\hline$\Delta \boldsymbol{\eta}$ & \multicolumn{2}{|c|}{12.0} & \multicolumn{2}{|c|}{10.5} & \multicolumn{2}{|c|}{13.7} \\
\hline$\Delta \phi$ & \multicolumn{2}{|c|}{14.3} & \multicolumn{2}{|c|}{23.5} & \multicolumn{2}{|c|}{24.5} \\
\hline
\end{tabular}
\end{table}

\begin{table} [h!]
\centering
\caption{The $\chi^2$ distances between the distributions of various jet variables obtained from Geant4, Delphes, and CALPAGAN for Dijet events.}
\label{table_2}
\begin{tabular}[H]{|c|c|c|c|c|c|c|}
\hline
\hline & \multicolumn{6}{|c|}{$\chi^2$ distances } \\
\hline & \multicolumn{2}{|c|}{ Delphes-Geant4 } & \multicolumn{2}{|c|}{ Geant4-CALPAGAN } & \multicolumn{2}{|c|}{ Delphes-CALPAGAN } \\
\hline & Jet 1 & Jet 2 & Jet 1 & Jet 2 & Jet 1 & Jet 2 \\
\hline 1-subjettiness & 179.0 & 140.3 & 118.0 & 99.3 & 83.1 & 71.5 \\
\hline 2-subjettiness & 431.0 & 400.3 & 259.5 & 238.4 & 137.2 & 157.1 \\
\hline Jet girth & 190.6 & 174.7 & 80.7 & 89.0 & 93.7 & 77.7 \\
\hline 2 point moment & 245.6 & 207.4 & 102.6 & 88.0 & 114.5 & 80.4 \\
\hline Jet $\mathbf{p}_{\mathrm{T}}$ & 75.4 & 54.4 & 17.4 & 38.6 & 99.5 & 102.4 \\
\hline Jet mass & 65.2 & 58.3 & 62.5 & 52.4 & 73.0 & 86.9 \\
\hline Jet $\boldsymbol{\eta}$ & 10.4 & 18.1 & 11.7 & 17.8 & 5.3 & 2.3 \\
\hline Jet $\phi$ & 3.4 & 5.5 & 8.1 & 6.8 & 3.4 & 2.5 \\
\hline$\Delta \eta$ & \multicolumn{2}{|c|}{9.3} & \multicolumn{2}{|c|}{17.3} & \multicolumn{2}{|c|}{16.1} \\
\hline$\Delta \phi$ & \multicolumn{2}{|c|}{24.8} & \multicolumn{2}{|c|}{15.7} & \multicolumn{2}{|c|}{11.5} \\
\hline
\end{tabular}
\end{table}

Table \ref{table_2} displays the $\chi^2$ distances between the distributions of various jet variables obtained from Geant4, Delphes, and CALPAGAN for Dijet events. 

Evaluating the speed enhancement of CALPAGAN in comparison to Geant4 was one of our primary objectives. Although given similar acceleration rates in similar studies, to make this comparison directly, the proposed model should provide all the outputs offered by the Geant4 simulation. Therefore, instead of comparing the inference time required for an event of the model , it was obtained for different batch size values for CPU, single GPU, and dual GPU. For the dual GPU setup, we used a model parallelism approach implemented in PyTorch. 

\begin{figure}[h!]
    \centering
    \includegraphics[width=\linewidth]{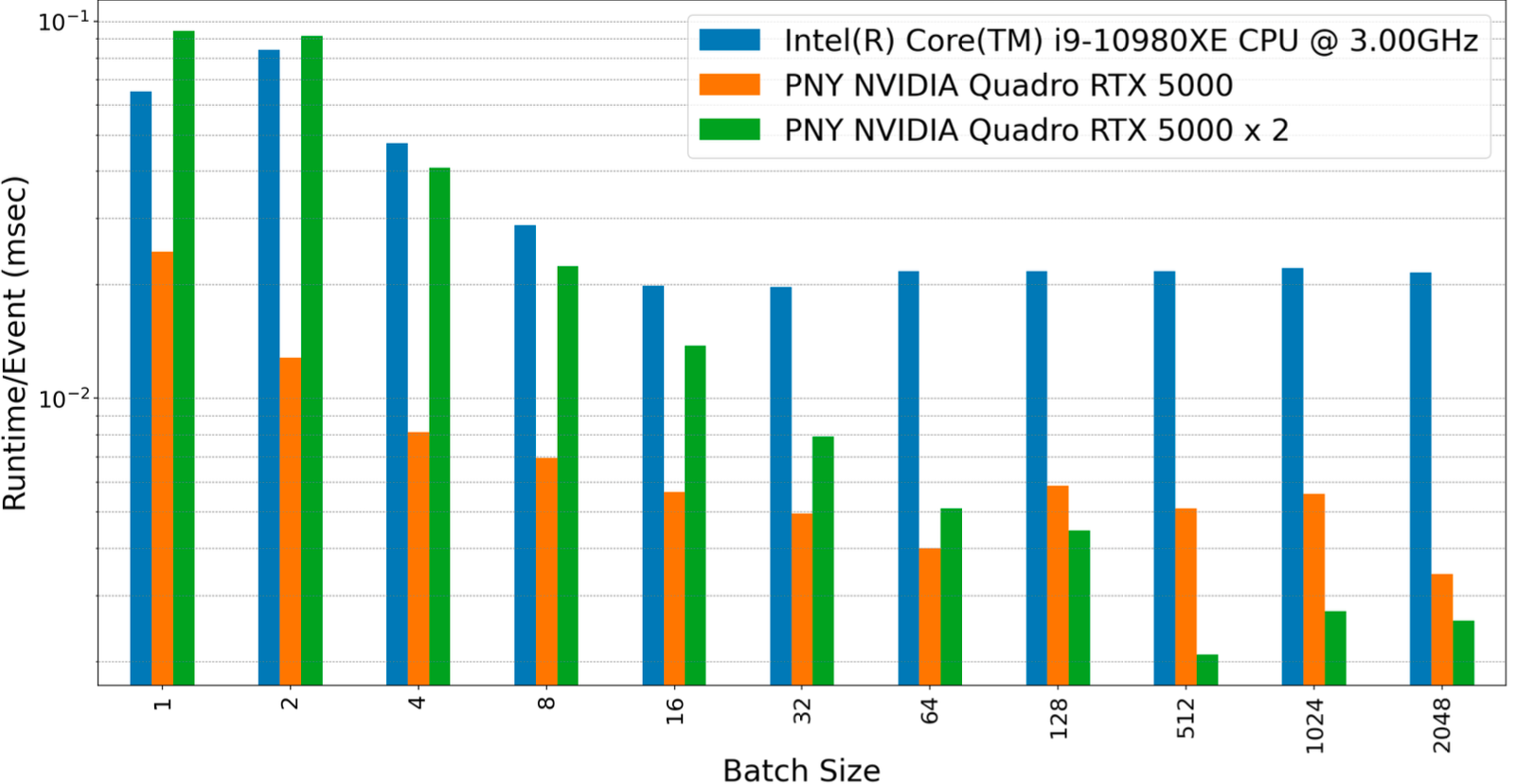}
    \caption{The inference time per event for CPU, single GPU, and dual GPU.}
    \label{fig:inference_time}
\end{figure}

As it can be seen in Fig.\ref{fig:inference_time}, GPU usage becomes more advantageous than CPU as the batch size increases. Of course, it is also important to note that the upper limit of batch size is RAM size (some may use swap memory with GPU, but in practice, this will increase inference time even more).

Moreover, as seen in \ref{fig:inference_time}, the dual GPU usage becomes advantageous for batch sizes 128 and more. This effect is most probably due to the data transfer latency. When using multiple GPUs, data needs to be transferred between them. For smaller batch sizes, the time taken to transfer data can overshadow the benefits of parallel computation. According to the results, the use of dual GPU makes the fastest inference per event when the batch size value is 512.
  \newpage

While evaluating CALPAGAN's performance in obtaining images, it has been promising to observe that the distributions of generative network results for various jet variables are closer to those obtained from Geant4 compared to Delphes. Furthermore, these images have been obtained much more rapidly compared to Geant4. However, this closeness applies only to a specific physical process during the training The training performed with W+Jet events has yielded the desired results for the W+Jet test sample, while the training with dijet events has provided the desired results for the dijet test sample. 

\section*{Acknowledgements}
This study was funded by the Scientific and Technological Research Council of Turkey (TUBITAK) ARDEB 1001 Grant No  119F084.

\newpage
\vspace{0.2cm}
\noindent
\let\doi\relax

{}
\end{document}